\documentclass[reprint,superscriptaddress,amsmath,amssymb,aps,pre]{revtex4-1}

\usepackage{amsmath}
\usepackage{graphicx}% Include figure files
\usepackage{dcolumn}% Align table columns on decimal point
\usepackage{bm}% bold math
\usepackage{color}
\usepackage{bbold}
\usepackage{soul}
\usepackage{subfigure}
\newcommand{\newc}{\newcommand}
\newc{\beq}{\begin{equation}}
\newc{\eeq}{\end{equation}}
\newc{\kt}{\rangle}
\newc{\br}{\langle}
\newc{\beqa}{\begin{eqnarray}}
\newc{\eeqa}{\end{eqnarray}}
\newc{\longra}{\longrightarrow}

\makeatletter
\let\Hy@backout\@gobble
\makeatother

\begin{document}

\title{An asymptotic relationship between homoclinic points and periodic orbit stability exponents}

\author{Jizhou Li}
\affiliation{Department of Physics and Astronomy, Washington State University, Pullman, Washington 99164-2814, USA}
\author{Steven Tomsovic}
\affiliation{Department of Physics and Astronomy, Washington State University, Pullman, Washington 99164-2814, USA}

\date{\today}

\begin{abstract}
The magnitudes of the terms in periodic orbit semiclassical trace formulas are determined by the orbits' stability exponents.  In this paper, we demonstrate a simple asymptotic relationship between those stability exponents and the phase-space positions of particular homoclinic points.  
\end{abstract}


\maketitle

\section{Introduction}
\label{Introduction}

A variety of properties of chaotic quantum systems can be calculated with semiclassical trace formulas, which are sums over certain sets of classical orbits (periodic, heteroclinic, or closed orbits, etc.) arising in their classical counterparts. For example, Gutzwiller's trace formula~\cite{Gutzwiller71} is a sum over periodic orbits that determines the spectrum, and the near-threshold absorption spectra of an atom in a magnetic field involve a sum over closed orbits that begin and end at the nucleus~\cite{Du88a,Du88b}.  Such orbit sums properly account for quantum interferences as each orbit carries a magnitude determined by its stability exponent, and a phase factor determined by its classical action and Maslov index. The orbits with shorter periods give long-range structure to the quantum spectra, and the longer period orbits the finer scale structure.  Due to exponential proliferation and instability, explicit construction of a complete set of orbits with longer and longer periods rapidly becomes prohibitive. 

In two previous publications~\cite{Li17a,Li18}, we developed an analytic scheme to express the classical actions of unstable periodic orbits in terms of action differences between certain homoclinic orbits. The homoclinic orbit action differences can then be obtained as phase-space integrals along the stable and unstable manifolds, which can be calculated stably by efficient numerical techniques~\cite{Li17,Krauskopf98b,Mancho03,Krauskopf05}.  Thus, the phase factors can be obtained via stable computations without the explicit construction of the orbits.  Here, we address the magnitudes, i.e.~the stability exponents of the periodic orbits.  A new relationship is developed to link the stability exponents of unstable periodic orbits to the phase-space positions of specific homoclinic points.  The exponent is determined by the ratio between relative positions from an asymptotic family of homoclinic points. Thus, the periodic orbit magnitudes can also be determined without explitic construction.  This implies a unified scheme of interchanging the periodic orbits with homoclinic orbits, which may be very benefiicial depending on the circumstances. 

The paper is organized as follows: Sec.~\ref{Basic Concepts and Definitions} introduces the basic concepts of hyperbolic orbits and the main language for the description of unstable orbits---symbolic dynamics.  A generic model for the symbolic dynamics, the Smale horseshoe~\cite{Smale63,Smale80}, is also introduced in this section. Sec.~\ref{Periodic orbit exponent} is the main content of this work that develops the central theorem. Sec.~\ref{Numerical verfication} provides numerical verification.  Sec.~\ref{Conclusion} makes a brief conclusion and points to directions for future work. 

\section{Basic Concepts}
\label{Basic Concepts and Definitions}

\subsection{Symbolic dynamics}
\label{Symbolic dynamics}

Let us consider a two-degree-of-freedom chaotic Hamiltonian system. With energy conservation and applying the Poincar\'{e} surface of section technique \cite{Poincare99}, the Hamiltonian flow is reduced to a discrete area-preserving map $M$ on the two-dimensional phase space $(q,p)$. Assuming the dynamics of the Hamiltonian systems is hyperbolic, the corresponding Poincar\'{e} map $M$ is also hyperbolic.  The orbit of a phase space point $z_0$, denoted by $\lbrace z_0 \rbrace$, is the bi-infinite collection of all $M^{n}(z_0)$: 
\begin{equation}\label{eq:Orbit}
\begin{split}
\lbrace z_0 \rbrace&=\lbrace \cdots,M^{-1}(z_0),z_0,M(z_0),\cdots \rbrace \\
&=\lbrace \cdots,z_{-1},z_0,z_1,\cdots \rbrace  \nonumber
\end{split}
\end{equation}
where $z_n = M^n(z_0)$ for all $n$. Generic orbits are hyperbolic, or exponentially unstable, as two orbits starting from nearby initial conditions will typically be separated exponentially under successive iterations. The exponential rate of a typical orbit is captured by the Lyapunov exponent, $\mu$, which quantifies the mean stretching and compressing rate of the hyperbolic map. In open systems, such stretching and compressing behaviors of the dynamics leads to certain $\mathit{escaping}$ $\mathit{orbits}$ that tend to infinity under successive inverse or forward iterations.  However, we concentrate on the orbits that do not escape to infinity, the $\mathit{nonwandering}$ $\mathit{set}$ (the results apply equally well in closed systems).  Denote the set of interest by $\Omega$. The main object of study in this article, namely the homoclinic and periodic orbits, all belong to $\Omega$. 

Let $x=(q,p)$ be a hyperbolic fixed point from $\Omega$, i.e., $M(x)=x$. Denote the unstable and stable manifolds of $x$ by $U(x)$ and $S(x)$, respectively. Typically, $U(x)$ and $S(x)$ intersect infinitely many times and form a complicated pattern called a homoclinic tangle~\cite{Poincare99,Easton86,Rom-Kedar90}.  The notation $U[a,b]$ is introduced to denote the finite segment of $U(x)$ extending from $a$ to $b$, both of which are points on $U(x)$, and similarly for $S(x)$. These manifolds are important skeleton-like structures of the dynamics since the exponential stretching and compressing of the map are fully captured by the unstable and stable manifolds, respectively. Furthermore, the folding of phase-space regions are also described by the folding of the manifolds. 

It is well-known that Markov partitions to the phase space \cite{Bowen75,Gaspard98} exist that use segments on $U(x)$ and $S(x)$ as boundaries, which are used to assign symbolic dynamics \cite{Hadamard1898,Birkhoff27a,Birkhoff35,Morse38} as phase-space itineraries of orbits in $\Omega$. The cells of the partition $\mathcal{V} = [V_0, V_1, \cdots, V_L]$ are closed curvilinear parallelograms bounded by the stable and unstable manifolds. The symbolic dynamics assigns a one-to-one correspondence between orbits of the system and sequences of symbols taken from an alphabet, $s_i \in [0,1,\cdots,L]$, which are in one-to-one correspondence with the cells $[V_0, V_1, \cdots, V_L]$ \cite{Gaspard98}. The symbolic code of a phase-space point $z_0$ is then a bi-infinite sequence of alphabets
\begin{equation}\label{eq:symbolic code}
z_0 \Rightarrow \cdots s_{-2}s_{-1} \cdot s_{0}s_{1}s_{2}\cdots  
\end{equation}
where each digit $s_n$ in the symbol denotes the cell that $M^{n}(z_0)$ belongs to: $M^{n}(z_0) = z_n \in V_{s_n}$, $s_n \in \lbrace 0,\cdots, L \rbrace$. The dot in the middle indicates the current iteration: $z_0 \in V_{s_0}$. In that sense, the symbolic code gives an ``itinerary" of $z_0$ under successive forward and backward iterations, in terms of the Markov cells in which each iteration lies. The mapping $M$ under the symbolic dynamics is then reduced to a simple shift of the dot in the code:
\begin{equation}\label{eq:Shift map}
M^n(z_0)=z_n \Rightarrow \cdots s_{n-1} \cdot s_{n}s_{n+1} \cdots  \nonumber.
\end{equation}
Points along the same orbit have the same symbolic strings but shifting dots. Therefore, an orbit can be represented by the symbolic string without the dot.  

\subsection{The Horseshoe map}
\label{The Horseshoe map}

\begin{figure}[ht]
\centering
{\includegraphics[width=7cm]{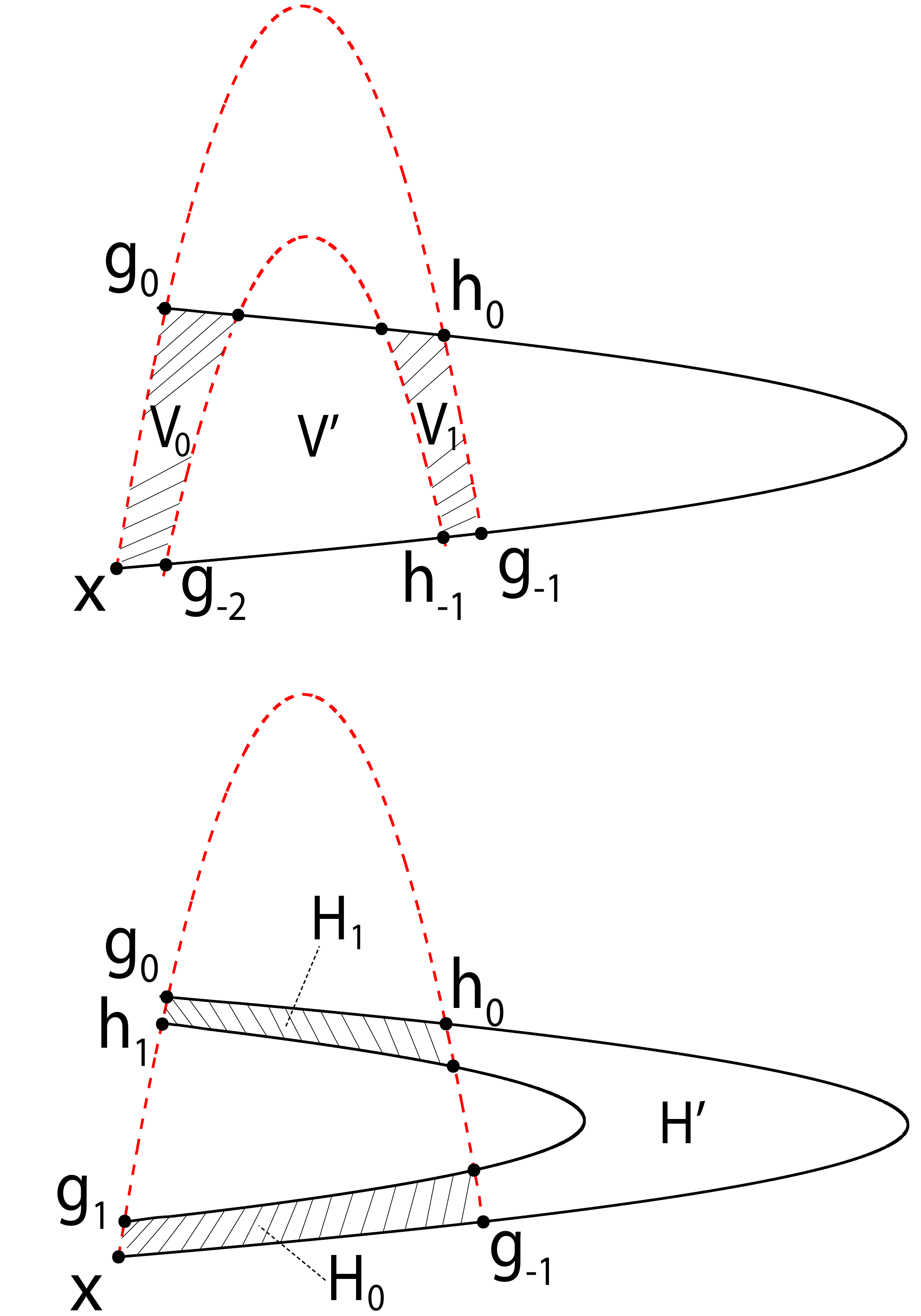}}
 \caption{Example partial homoclinic tangle from the H\'{e}non map, which forms a complete horseshoe structure. The unstable (stable) manifold of $x$ is the solid (dashed) curve. There are two primary homoclinic orbits $\lbrace h_0\rbrace$ and $\lbrace g_0\rbrace$. Let $\cal{R}$ be the closed region bounded by manifold segments $U[x,g_{-1}]$, $S[g_{-1},h_0]$, $U[h_0,g_0]$, and $S[g_0,x]$. In the upper panel, $\cal{R}$ can be identified as the region composed by $V_0$, $V^{\prime}$, and $V_1$. Under forward iteration, the vertical strips $V_0$ and $V_1$ (including the boundaries) from the upper panel are mapped into the horizontal strips $H_0$ and $H_1$ in the lower panel. At the same time, points in region $V^{\prime}$ are mapped outside $\cal{R}$ into region $H^{\prime}$, never to return and escape to infinity. There is a Cantor set of points in $V_0$ and $V_1$ that remain inside $\cal{R}$ for all iterations, which is the non-wandering set $\Omega$. The phase-space itineraries of points in $\Omega$ in terms of $V_0$ and $V_1$ give rise to symbolic dynamics.}
\label{fig:horseshoe}
\end{figure}

Assuming the system is highly chaotic, so the homoclinic tangle forms a complete horseshoe, part of which is shown in Fig.~\ref{fig:horseshoe}, as this is generic to a significant class of
dynamical systems. In such scenarios, the Markov partition is a simple set of two regions $[V_0,V_1]$, as shown in the upper panel of Fig.~\ref{fig:horseshoe}. Each phase-space point $z_0$ that never escapes to infinity can be put into an one-to-one correspondence with a bi-infinite symbolic string in Eq.~\eqref{eq:symbolic code}, where each digit $s_{n} \in {0,1}$ such that $M^{n}(z_0) \in V_{s_n}$. 

Throughout this paper we use the area-preserving H\'{e}non map \cite{Henon76} with parameter $a=10$ for illustration and numerical implementations:
\begin{equation}\label{eq:Henon map}
\begin{split}
&p_{n+1}=q_n\\
&q_{n+1}=a-q_{n}^2-p_n.
\end{split}
\end{equation}

This parameter is well beyond the first tangency, thus giving rise to a complete horseshoe-shaped homoclinic tangle with highly chaotic dynamics. It serves as a simple paradigm since the symbolic dynamics permits all possible combinations of binary codes, no ``pruning''~\cite{Cvitanovic88a,Cvitanovic91} is needed.  The results derived ahead mostly carry over into more complicated systems possessing incomplete horseshoes, or systems with more than binary symbolic codes, though more work is needed to address such systems.  Appendix A of~\cite{Li18} has more details on the partition and symbolic dynamics relevant here. 

The intersections between $S(x)$ and $U(x)$ give rise to homoclinic orbits, which are asymptotic to $x$ under both $M^{\pm\infty}$. From the infinite families of homoclinic orbits, two special ones $\lbrace h_0\rbrace$ and $\lbrace g_0\rbrace$ can be identified as primary homoclinic orbits, in the sense that they have the simplest phase space excursions. The segments $S[x,h_0]$ and $U[x,h_0]$ intersect only at $h_0$ and $x$, the same is true for all its orbit points $h_i$; this holds for $\lbrace g_0\rbrace$ as well. There are only two primary orbits for the horseshoe, but possibly more for systems with more complicated homoclinic tangles.   

Under the symbolic dynamics, a period-$T$ point $y_0$, where $M^T(y_0)=y_0$, can always be associated with a symbolic string with infinite repetitions of a substring with length $T$: 
\begin{equation}\label{eq:Periodic point}
 y_0  \Rightarrow \cdots s_0 s_1 \cdots s_{T-1} \cdot s_0 s_1 \cdots s_{T-1} \cdots =\overline{\gamma}.\overline{\gamma}
\end{equation}
where $\gamma =s_0 \cdots s_{T-1}$ is the finite substring and $\overline{\gamma}.\overline{\gamma}$ denotes its infinite repetition (on both sides of the dot). Notice that the cyclic permutations of $s_0 \cdots s_{T-1}$ can be associated with the successive mappings of $y_0$, generating a one-to-one mapping to the set of points on the orbit. Since an orbit can be represented by any point on it, the position of the dot does not matter, therefore we denote the periodic orbit $\lbrace y_0 \rbrace$ as
\begin{equation}\label{eq:Periodic orbit}
\lbrace  y_0 \rbrace \Rightarrow \overline{\gamma} 
\end{equation}  
with the dot removed.  Similarly, the finite length-$T$ orbit segment $[y_0,y_1,\cdots,y_{T-1}]$, which composes one full period,  is denoted
\begin{equation}\label{eq:Periodic orbit segment}
{y_0,y_1,\cdots,y_{T-1}} \Rightarrow \gamma 
\end{equation}  
with the overhead bar removed, as compared to Eq.~\eqref{eq:Periodic orbit}.  Any cyclic permutation of $\gamma$ refers to the same periodic orbit.
 
The hyperbolic fixed point has the simplest symbolic code $x \Rightarrow \overline{0} \cdot \overline{0}$\ , and its orbit $\lbrace x \rbrace \Rightarrow \overline{0}$ correspondingly. A homoclinic point $h_0$ of $x$ has symbolic code of the form~\cite{Hagiwara04}: 
\begin{equation}\label{eq:Homoclinic point}
h_0 \Rightarrow \overline{0} 1 s_{-m}\cdots s_{-1} \cdot s_0 s_1 \cdots s_n 1 \overline{0} 
\end{equation}
along with all possible shifts of the dot, where the $\overline{0}$ on both ends means the orbit approaches the fixed point (therefore stays in $V_0$) under both $M^{\pm \infty}$. Similar to the periodic orbit case, the homoclinic orbit can be represented as
\begin{equation}\label{eq:Homoclinic orbit}
\lbrace h_0 \rbrace \Rightarrow  \overline{0} 1 s_{-m}\cdots s_{-1}s_0 s_1 \cdots s_n 1 \overline{0}
\end{equation}
with the dot removed, as compared to Eq.~\eqref{eq:Homoclinic point}. 

A \textit{heteroclinic} orbit $\lbrace h^{\prime}_0 \rbrace$ between the periodic point $y \Rightarrow \overline{\gamma} \cdot \overline{\gamma}$ and  the fixed-point $x \Rightarrow \overline{0} \cdot \overline{0}$ arises from $h^{\prime}_0 = U(y) \cap S(x)$, and can be represented by
\begin{equation}\label{eq:Heteroclinic orbit symbolic code}
\lbrace h^{\prime}_0 \rbrace \Rightarrow \overline{\gamma} \gamma^{\prime} \overline{0}  
\end{equation}
where the asymptotic behaviors of the orbit are described by $\overline{\gamma}$ and $\overline{0}$ on the two ends, and the finite symbolic string $\gamma^{\prime}$ describes the connection from $\lbrace y \rbrace$ to $\lbrace x \rbrace$, which solely depends on the choice of $h^{\prime}_0$.

\section{Periodic orbit stability exponent}
\label{Periodic orbit exponent}

Consider an arbitrary unstable periodic orbit $\lbrace y  \rbrace$ with symbolic code $\lbrace y  \rbrace \Rightarrow \overline{\gamma}$. Let the length of the symbolic string $\gamma$ be $n_{\gamma}$, which is also the periodic of $\lbrace y \rbrace$. The periodic point $y$ can also be viewed as a fixed-point under the $n_{\gamma}$-th compound mapping of $M$: $M^{n_{\gamma}}(y) = y$. Denote the eigenvalue of the unstable subspace of the tangent space of $\lbrace y \rbrace$ under one full period ($n_{\gamma}$) by $\lambda_{\gamma}$. Thus $\lambda_{\gamma}> 1$ if $\lbrace y \rbrace$ is hyperbolic without reflection ($\lambda_{\gamma} <-1$ with reflection). The stability exponent of $\lbrace y \rbrace$, denoted by $\mu_{\gamma}$, is then $n_{\gamma}\mu_{\gamma} = \ln | \lambda_{\gamma}| $. 

To help determine $\lambda_\gamma$ and thus $\mu_{\gamma}$, choose a family of auxiliary homoclinic points of the fixed-point $x$, namely $h^{(m)}_{0}$ ($m =1,2,\cdots$), that has the symbolic codes
\begin{equation}
\label{eq:Auxiliary orbit symbolic codes}
h^{(m)}_0  \Rightarrow \overline{0}  \gamma^{m} \cdot  \overline{0} 
\end{equation}
where $\gamma^{m}$ denotes $m$ repetitions of $\gamma$ and $m=1,2,\cdots$.  Having identified the auxiliary homoclinic points, let us consider the homoclinic orbit segments generated by certain numbers of inverse iterations of them, namely
\begin{equation}
\label{eq:Auxiliary homoclinic orbit segments}
\textrm{Seg(k,m)} = \lbrace h^{(m)}_{-N(k,m)}, \cdots, h^{(m)}_{-1}, h^{(m)}_0 \rbrace
\end{equation}
where $N(k,m)=(k + m)n_{\gamma}$ is a positive integer determined by $k$ and $m$ ($k , m \geq 1$).  Ahead $k$ is taken to $\infty$, which yields the limit
\begin{equation}
\label{eq:Auxiliary orbit segments past asymptote}
\lim_{k \to \infty} h^{(m)}_{-N(k,m)} = x \ .
\end{equation}
  
The key to the derivation lies in the \textit{normal-form} transformation~\cite{Birkhoff27,Moser56,Silva87} of  three orbit segments. 
\begin{figure}[ht]
\centering
{\includegraphics[width=0.9\linewidth]{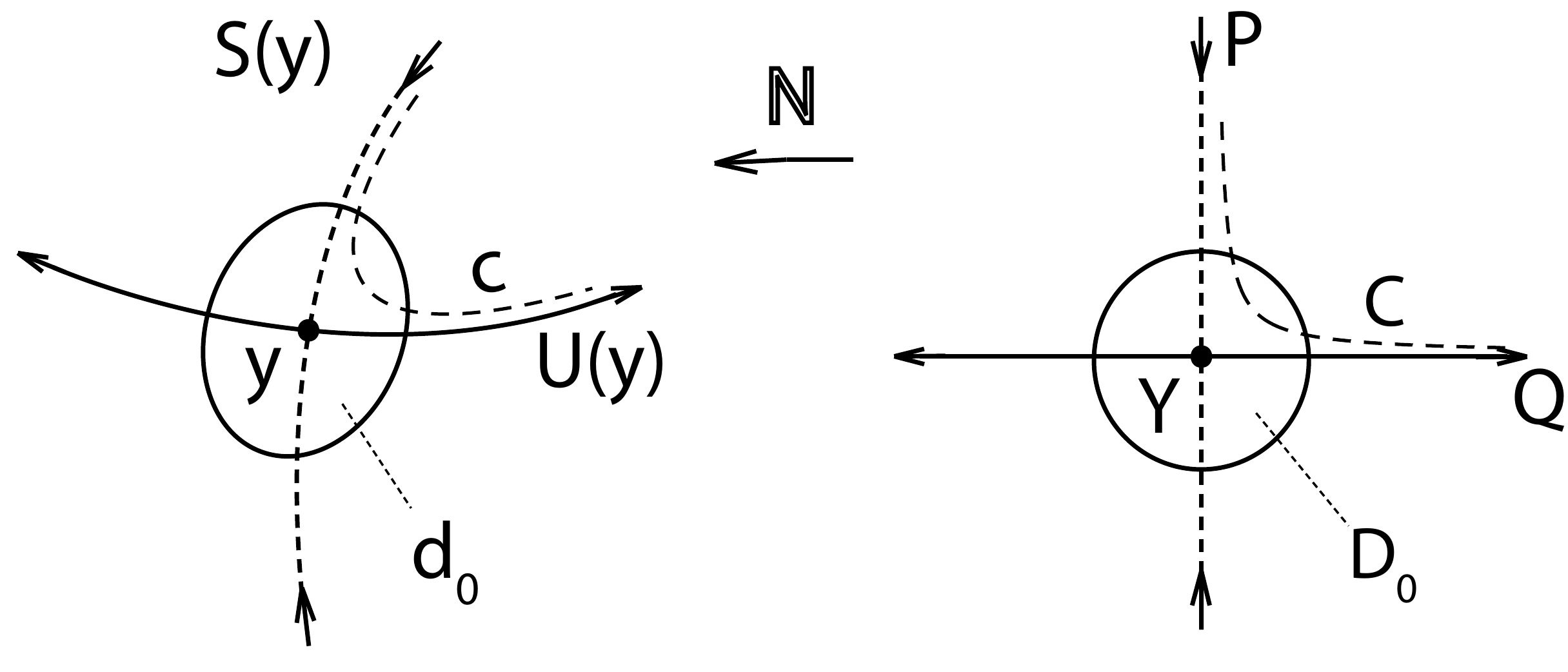}}
 \caption{Schematic visualization of the normal form transformation $\mathbb{N}$. It transforms points from the normal form coordinate $(Q,P)$ into the phase space coordinate $(q,p)$. The $Q$ and $P$ axis are mapped into $U(y)$ and $S(y)$, respectively. The advantage of normal form coordinates is that the dynamics preserves the $QP$ product [Eq.~\eqref{eq:normal form}], thus points are mapped along invariant hyperbolas, as shown by $C$ in the right panel. The family of invariant hyperbolas then give rise to a family of Moser invariant curves in phase space via the transformation $\mathbb{N}(C)=c$, shown in the left panel.}
\label{fig:Normal_form_Intro}
\end{figure}  
For well-behaved (invertible and analytic) Poincar\'{e} maps, the nonlinear dynamics near the stable and unstable manifolds can be linearized via a common technique called \textit{normal-form transformation}, denoted by $\mathbb{N}$, which transforms points from the normal form coordinates $(Q,P)$ to the neighborhood of stable and unstable manifolds of the hyperbolic fixed point $y$: $\mathbb{N}: (Q,P) \mapsto (q,p)$, as shown in Fig.~\ref{fig:Normal_form_Intro}. In the normal form coordinates of $y$, the compound mapping $M^{n_{\gamma}}$ takes a simple form:  
\begin{equation}\label{eq:normal form}
\begin{split}
&Q_{n+1}=\Lambda(Q_{n}P_{n})\cdot Q_{n}\\
&P_{n+1}=[\Lambda(Q_{n}P_{n})]^{-1}\cdot P_{n}
\end{split}
\end{equation}
where $\Lambda(Q_{n}P_{n})$ is a polynomial function of the product $Q_{n}P_{n}$ \cite{Harsoula15}: 
\begin{equation}\label{eq:Lambda}
\Lambda(QP)=\lambda_{\gamma}+w_{2}\cdot (QP)+w_{3}\cdot (QP)^{2}+\cdots
\end{equation}
The normal form convergence zone was first proved by Moser~\cite{Moser56} to be a small disk-shaped region centered at the fixed point ($D_0$ and its image $d_0$ in Fig.~\ref{fig:Normal_form_Intro}), and later proved by da Silva Ritter $\mathit{et.}$ $\mathit{al.}$~\cite{Silva87} to extend along the stable and unstable manifolds to infinity. The extended convergence zone follows hyperbolas to the manifolds  (``gets exponentially close'' the further out along the manifolds). The stable and unstable manifolds are just images of the $P$ and $Q$ axes respectively under the normal form transformation.  

All points inside the extended convergence zone near the $Q$ or $P$ axis move along invariant hyperbolas, which are mapped to Moser invariant curves in phase space.  A schematic example is shown in Fig.~\ref{fig:Normal_form_Intro}, where the hyperbola $C$ in the normal form coordinates is transformed into a Moser curve $c$ in phase space. Being confined in the extended convergence zone, the Moser invariant curves also get exponentially close to the stable and unstable manifolds while extending along them outward to infinity. In fact, as shown by~\cite{Harsoula15}, the convergence zone can be quantified using the outermost Moser curve with the largest $QP$ product.  

\begin{figure}[ht]
\centering
{\includegraphics[width=7cm]{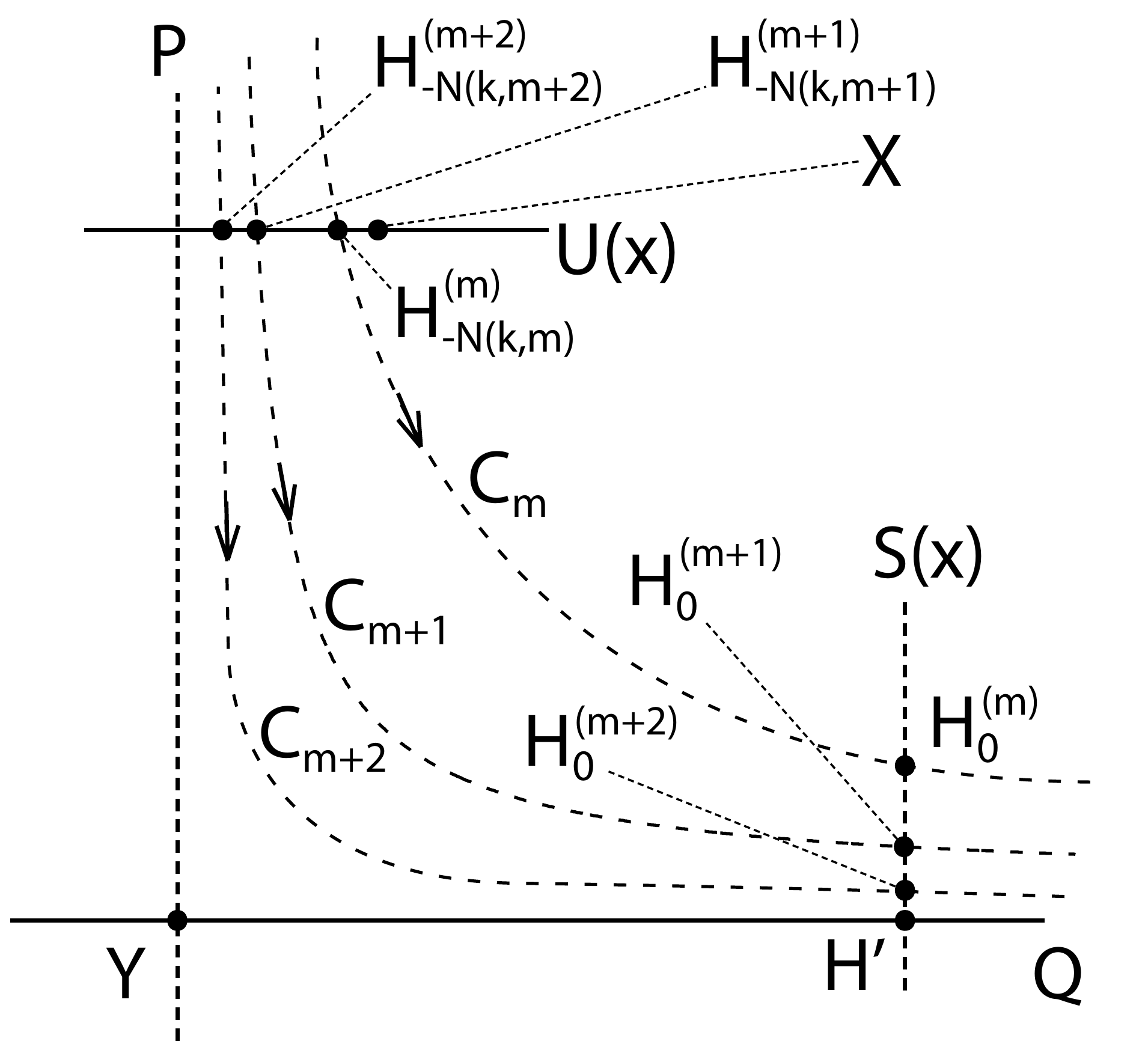}}
 \caption{(Schematic) Normal-form coordinate picture of the auxiliary homoclinic orbit segments. $Y$ is the image of $y$, and the $P$, $Q$ axis are the images of $S(y)$ and $U(y)$, respectively, in the normal-form coordinate. $X$ is the image of the fixed-point $x$. Three auxiliary homoclinic orbit segments, corresponding to $m$, $m+1$, and $m+2$ in Eq.~\eqref{eq:Auxiliary homoclinic orbit segments normal form}, lie on the hyperbolas $C_m$, $C_{m+1}$, and $C_{m+2}$, respectively. Note that only the first and the last points of each orbit segment are drawn here.    }
\label{fig:Normal_form_orbits}
\end{figure}     

Let the image of Seg$(k,m)$ in the normal-form coordinate of $y$ be
\begin{equation}
\label{eq:Auxiliary homoclinic orbit segments normal form}
\textrm{Seg(k,m)} = \lbrace H^{(m)}_{-N(k,m)}, \cdots, H^{(m)}_{-1}, H^{(m)}_0 \rbrace
\end{equation}
where $\mathbb{N}(H^{(m)}_n) = h^{(m)}_n$. In the normal-form coordinates, every Seg$(k,m)$ lies on a hyperbola, labeled by $C_m$ in Fig.~\ref{fig:Normal_form_orbits}. This figure shows Seg$(k,m)$, Seg$(k,m+1)$, and Seg$(k,m+2)$, in the normal-form coordinate of the periodic point $y$. Letting $k \to \infty$, because of Eq.~\eqref{eq:Auxiliary orbit segments past asymptote}, the initial points of the three segments, namely $H^{(m)}_{-N(k,m)}$, $H^{(m+1)}_{-N(k,m+1)}$, and $H^{(m+2)}_{-N(k,m+2)}$, are all located infinitesimally close to $X$ along $U(x)$:
\begin{equation}\label{eq:Auxiliary orbit segments past asymptote normal form}
\begin{split}
&\lim_{k\to\infty} H^{(m)}_{-N(k,m)} = \lim_{k\to\infty} H^{(m+1)}_{-N(k,m+1)} \\
& = \lim_{k\to\infty} H^{(m+2)}_{-N(k,m+2)} = X\ .
\end{split}
\end{equation}

Then, under $N(k,m)$, $N(k,m+1)$, and $N(k,m+2)$ iterations, respectively, they are mapped to the final points $H^{(m)}_0$, $H^{(m+1)}_0$, and $H^{(m+2)}_0$, as shown near the heteroclinic point $H^{\prime}$ on the $Q$ axis. This heteroclinic point has symbolic code
\begin{equation}\label{eq:Heteroclinic point h prime}
\mathbb{N}(H^{\prime}) = h^{\prime} \Rightarrow \overline{\gamma} \cdot \overline{0}
\end{equation}
which is the simplest heteroclinic connection between $\lbrace y \rbrace$ and $\lbrace x \rbrace$.  Recall that
\begin{equation}\label{eq:Hm symbolic code}
\mathbb{N}(H^{(m)}_0) = h^{(m)}_0 \Rightarrow \overline{0} \gamma^{m} \cdot \overline{0}\ .
\end{equation}
Comparing the symbolic strings in Eqs.~\eqref{eq:Heteroclinic point h prime} and \eqref{eq:Hm symbolic code}, it follows that the codes of $h^{\prime}$ and $h^{(m)}_0$ to the right of the dot are identical, and to the left  of the dot they match up to $\gamma^{m}$ (which has length $mn_{\gamma}$). This indicates that $h^{(m)}_0$ is $\sim O(e^{-mn_{\gamma} \mu_\gamma})$ close to $h^{\prime}$ along $S(x)$ (for more details, see Appendix A of Ref.~\cite{Li18}). Due to the same reason, $h^{(m+1)}_0$ and $h^{(m+2)}_0$ are $\sim O(e^{-(m+1)n_{\gamma} \mu_\gamma})$ and $\sim O(e^{-(m+2)n_{\gamma} \mu_\gamma})$ close to $h^{\prime}$, respectively, along $S(x)$. Therefore, the above four points, $h^{\prime}$, $h^{(m)}_0$, $h^{(m+1)}_0$, and $h^{(m+2)}_0$, are all within exponentially small neighborhoods of each other. 

The same conclusion holds true in the normal-form coordinates. In fact, as shown by Fig.~\ref{fig:Normal_form_orbits}, in the normal coordinate of $y$, the proportionality factors of the distances between them can be determined analytically. Plotted in the figure are the initial and final points of Seg$(k,m)$, Seg$(k,m+1)$, and Seg$(k,m+2)$. The $k$ here is assumed to be a large integer, so $H^{(m)}_{-N(k,m)}$, $H^{(m+1)}_{-N(k,m+1)}$, and $H^{(m+2)}_{-N(k,m+2)}$ are exponentially close to $X$. Under successive forward iterations, they are mapped along the hyperbolas $C_{m}$, $C_{m+1}$, and $C_{m+2}$, respectively, into $H^{(m)}_0$, $H^{(m+1)}_0$, and $H^{(m+2)}_0$:
\begin{equation}
M^{N(k,m+i)}( H^{(m+i)}_{-N(k,m+i)} ) = H^{(m+i)}_0
\end{equation}
where $i=0,1,2$. The mapping equations take the simple form of Eq.~\eqref{eq:normal form}, with the stability factor $\Lambda$ given by
\begin{equation}\label{eq:Normal form factor}
\Lambda(QP) = \lambda_\gamma + w_2 \cdot (QP) + w_3 \cdot (QP)^2 + \cdots\ .
\end{equation}
Under the limit $m \to \infty$, $H^{(m+i)}_0 \to H^{\prime}$ ($i=0,1,2$), so their $QP$ products along the hyperbolas $\to 0$. Correspondingly, the stability factor $\Lambda(QP) \to \lambda_\gamma$, and the $C_m$ curve becomes infinitely close to the $P$ and $Q$ axis when $m \to \infty$.

Consequently, the $P$ coordinate values of $H^{(m+i)}_0$, namely $P(H^{(m+i)}_0)$, are determined asymptotically by
\begin{equation}\label{eq:Y coordinate values of H}
 \lim_{m \to \infty} P(H^{(m+i)}_0) = \lim_{\substack{
   k \to \infty \\
   m \to \infty
  }} P(X) \cdot \lambda_{\gamma}^{-(k+m+i)}
\end{equation}
for $i=0,1,2,\cdots$, where $P(X)$ denotes the $P$-coordinate value of $X$. Furthermore, notice that $N(k,m+j)-N(k,m)=j \cdot n_{\gamma}$, thus using Eq.~\eqref{eq:Y coordinate values of H} we get
\begin{equation}\label{eq:Y coordinate scaling of H}
\lim_{m \to \infty} P(H^{(m+j)}_0) = \lim_{m \to \infty} P(H^{(m)}_0) \cdot \lambda_{\gamma}^{-j}
\end{equation}
for $j=0,1,2,\cdots$. Therefore, the family of homoclinic points, $H^{(m+j)}_{0}$ ($j=0,1,2,\cdots$), converges to $H^{\prime}$ under convergence factor $\lambda_{\gamma}^{-1}$. Since the normal-form transformation preserves the convergence factor of asymptotic series of points (see Appendix B.2 of Ref.~\cite{Mitchell03a} for a detailed proof), the family $h^{(m+j)}_{0}$ ($j=0,1,2,\cdots$) also converge to $h^{\prime}$ in phase space. Therefore, their phase-space positions satisfy
\begin{equation}\label{eq:Phase-space scaling of h}
\begin{split}
 \lim_{m \to \infty} \frac{ p(h^{(m)}_0) - p(h^{\prime}) } { p(h^{(m+j)}_0) - p(h^{\prime}) } = \lambda_{\gamma}^{j} \\
\lim_{m \to \infty} \frac{  q(h^{(m)}_0) - q(h^{\prime})  } { q(h^{(m+j)}_0) - q(h^{\prime}) } = \lambda_{\gamma}^{j}
\end{split}
\end{equation}
where $p(a)$ and $q(a)$ denotes the $p$- and $q$- coordinate values, respectively, of the point $a$. Here it is assumed the generic case that the local direction of $S(x)$ at $h^{\prime}$ is not strictly vertical or horizontal, so the differences between the $p$ and $q$ values of successive members do not vanish. The distances between successive members of the family are also in scale:
\begin{equation}\label{eq:Phase-space scaling of distances between h}
\lim_{m \to \infty} \frac{ p(h^{(m)}_0) - p(h^{(m+1)}_0 )  }{ p(h^{(m+1)}_0) - p(h^{(m+2)}_0) } = \lambda_{\gamma}
\end{equation}
and the same is true for the $q$ coordinate values as well. Therefore, the stability exponent $n_\gamma \mu_{\gamma} = \ln |\lambda_{\gamma}|$ of the periodic orbit $\lbrace y \rbrace \Rightarrow \overline{\gamma}$ can be determined using Eq.~\eqref{eq:Phase-space scaling of distances between h} from the family of auxiliary homoclinic points $h^{(m)}_0 \Rightarrow \overline{0} \gamma^{m} \cdot \overline{0}$, which does not require the numerical construction of the periodic orbit. In practice, for long periodic orbits with large periods ($n_{\gamma}$), the leading terms in the $h^{(m)}_0$ family should provide an accurately enough calculation of $\lambda_{\gamma}$:
\begin{equation}\label{eq:Phase-space scaling of distances between h approx}
\lambda_{\gamma} \approx \frac{ p(h^{(1)}_0) - p(h^{(2)}_0 )  }{ p(h^{(2)}_0) - p(h^{(3)}_0) }\ .
\end{equation}
 
\section{Explicit example}
\label{Numerical verfication}

To verify Eq.~\eqref{eq:Phase-space scaling of distances between h approx}, we have numerically constructed four different periodic orbits in the H\'{e}non map (Eq.~\eqref{eq:Henon map}), namely $\lbrace v \rbrace$, $\lbrace y \rbrace$, $\lbrace w \rbrace$, and $\lbrace z \rbrace$, with symbolic codes
\begin{equation}\label{eq:Numerics periodic orbit symbolic codes}
\begin{split}
\lbrace v \rbrace & \Rightarrow \overline{1011} \\
\lbrace y \rbrace & \Rightarrow \overline{0011} \\
\lbrace w \rbrace & \Rightarrow \overline{0001} \\
\lbrace z \rbrace & \Rightarrow \overline{00011}\ .
\end{split}
\end{equation}
The phase-space positions of one of their orbit points are 
\begin{equation}\label{eq:Numerics periodic orbit points}
\begin{split}
v & = ( 3.162277660168, 1.917144929227 ) \\
y & = (-3.162277660168, 3.162277660168 ) \\
w & = (-4.040365740912, -3.162277660168) \\
z &= ( -3.300504906006,  3.181101045340  )
\end{split}
\end{equation}
which are mapped back into themselves under their respective periods. Using these numerical orbits, their respective stability eigenvalues $\lambda_{\gamma}$ and exponents $\mu_{\gamma}$ have been calculated.  In addition, by constructing the respective auxiliary homoclinic points in Eq.~\eqref{eq:Auxiliary orbit symbolic codes} for each orbit, the same stability eigenvalues $\lambda^{\prime}_{\gamma}$ and exponents $\mu^{\prime}_{\gamma}$ have been approximated with Eq.~\eqref{eq:Phase-space scaling of distances between h approx}. The results are listed in Table \ref{tab:Numerical results}.  Although only using the leading terms in each auxiliary homoclinic family, the agreement is excellent. 
\begin{table}[h!]
  \begin{center}
    \begin{tabular}{l|c|c|c|c}
     $\overline{\gamma}$  & $\lambda_{\gamma}$ & $\lambda^{\prime}_{\gamma}$ & $n_\gamma \mu_{\gamma}$ & $n_\gamma \mu^{\prime}_{\gamma}$   \\
      \hline
      $\overline{1011}$ & $-586.069$  & $-584.741$  &  $6.37343$ & $6.37116$  \\
      $\overline{0011}$ & $1602.00$ &  $1602.20$  &  $7.37900$ & $7.37913$  \\
      $\overline{0001}$ & $-2609.92$ &  $-2609.72$  &  $7.86707$  & $7.86699$ \\
      $\overline{00011}$ & $14176.1$  &  $14180.5$  &  $9.55931$ & $9.55962$ \\
    \end{tabular}
    \caption{Unstable eigenvalues and the corresponding exponents of the periodic orbits in Eq.~\eqref{eq:Numerics periodic orbit symbolic codes}. $\lambda_{\gamma}$ are calculated from the numerical orbits, and $\lambda^{\prime}_{\gamma}$ are determined from Eq.~\eqref{eq:Phase-space scaling of distances between h approx}. The exponents are obtained as $n_\gamma \mu_{\gamma} = \ln|\lambda_{\gamma}|$ and $n_\gamma\mu^{\prime}_{\gamma} = \ln|\lambda^{\prime}_{\gamma}|$. }
    \label{tab:Numerical results}
  \end{center}
\end{table}

\section{Conclusion}
\label{Conclusion}

An exact formula (Eq.~\eqref{eq:Phase-space scaling of distances between h}) is introduced that links the stability properties of unstable periodic orbits to the phase space locations of certain homoclinic points. Although the formula is asymptotic in nature, the numerical results from using the leading term already reproduces the actual exponents quite accurately in the numerical model used.  Since the numerical computation of long periodic orbits suffers from an exponential instability problem, whereas the positions of homoclinic points can be determined relatively easily as the intersections between the invariant manifolds~\cite{Li17}, this approach may provide an efficient alternative to direct calculations. Furthermore, in previous work~\cite{Li17a,Li18}, the classical actions of periodic orbits are expressed in terms of certain homoclinic orbit action differences. Combined with the current results, they provide a unified scheme of replacing the periodic orbits in the trace formula by homoclinic orbits, which may lead to new resummation techniques in semiclassical methods. 

An important generalization of the current theory would be to extend it to higher-dimensional symplectic maps with chaotic dynamics.  For instance, in $4$-dimensional maps, the stable and unstable manifolds of hyperbolic fixed points will each be $2$-dimensional surfaces. They intersect in the $4$D phase-space generating homoclinic points. It may be the case that the relative positions of certain homoclinic points distributed along the dominant contraction direction in the stable manifolds yield the dominant stability exponent, and the relative positions of the homoclinic points along the sub-dominant contraction direction of the stable manifolds yield the sub-dominant stability exponent. However, the generalization of the symbolic code description to higher-dimensions is a challenging issue. 

\bibliography{classicalchaos,quantumchaos}

%merlin.mbs apsrev4-1.bst 2010-07-25 4.21a (PWD, AO, DPC) hacked
%Control: key (0)
%Control: author (8) initials jnrlst
%Control: editor formatted (1) identically to author
%Control: production of article title (-1) disabled
%Control: page (0) single
%Control: year (1) truncated
%Control: production of eprint (0) enabled
\begin{thebibliography}{29}%
\makeatletter
\providecommand \@ifxundefined [1]{%
 \@ifx{#1\undefined}
}%
\providecommand \@ifnum [1]{%
 \ifnum #1\expandafter \@firstoftwo
 \else \expandafter \@secondoftwo
 \fi
}%
\providecommand \@ifx [1]{%
 \ifx #1\expandafter \@firstoftwo
 \else \expandafter \@secondoftwo
 \fi
}%
\providecommand \natexlab [1]{#1}%
\providecommand \enquote  [1]{``#1''}%
\providecommand \bibnamefont  [1]{#1}%
\providecommand \bibfnamefont [1]{#1}%
\providecommand \citenamefont [1]{#1}%
\providecommand \href@noop [0]{\@secondoftwo}%
\providecommand \href [0]{\begingroup \@sanitize@url \@href}%
\providecommand \@href[1]{\@@startlink{#1}\@@href}%
\providecommand \@@href[1]{\endgroup#1\@@endlink}%
\providecommand \@sanitize@url [0]{\catcode `\\12\catcode `\$12\catcode
  `\&12\catcode `\#12\catcode `\^12\catcode `\_12\catcode `\%12\relax}%
\providecommand \@@startlink[1]{}%
\providecommand \@@endlink[0]{}%
\providecommand \url  [0]{\begingroup\@sanitize@url \@url }%
\providecommand \@url [1]{\endgroup\@href {#1}{\urlprefix }}%
\providecommand \urlprefix  [0]{URL }%
\providecommand \Eprint [0]{\href }%
\providecommand \doibase [0]{http://dx.doi.org/}%
\providecommand \selectlanguage [0]{\@gobble}%
\providecommand \bibinfo  [0]{\@secondoftwo}%
\providecommand \bibfield  [0]{\@secondoftwo}%
\providecommand \translation [1]{[#1]}%
\providecommand \BibitemOpen [0]{}%
\providecommand \bibitemStop [0]{}%
\providecommand \bibitemNoStop [0]{.\EOS\space}%
\providecommand \EOS [0]{\spacefactor3000\relax}%
\providecommand \BibitemShut  [1]{\csname bibitem#1\endcsname}%
\let\auto@bib@innerbib\@empty
%</preamble>
\bibitem [{\citenamefont {Gutzwiller}(1971)}]{Gutzwiller71}%
  \BibitemOpen
  \bibfield  {author} {\bibinfo {author} {\bibfnamefont {M.~C.}\ \bibnamefont
  {Gutzwiller}},\ }\href@noop {} {\bibfield  {journal} {\bibinfo  {journal}
  {J.~Math.~Phys.}\ }\textbf {\bibinfo {volume} {12}},\ \bibinfo {pages} {343}
  (\bibinfo {year} {1971})},\ \bibinfo {note} {and references
  therein}\BibitemShut {NoStop}%
\bibitem [{\citenamefont {Du}\ and\ \citenamefont
  {Delos}(1988{\natexlab{a}})}]{Du88a}%
  \BibitemOpen
  \bibfield  {author} {\bibinfo {author} {\bibfnamefont {M.~L.}\ \bibnamefont
  {Du}}\ and\ \bibinfo {author} {\bibfnamefont {J.~B.}\ \bibnamefont {Delos}},\
  }\href@noop {} {\bibfield  {journal} {\bibinfo  {journal} {Phys.~Rev.~A}\
  }\textbf {\bibinfo {volume} {38}},\ \bibinfo {pages} {1896} (\bibinfo {year}
  {1988}{\natexlab{a}})}\BibitemShut {NoStop}%
\bibitem [{\citenamefont {Du}\ and\ \citenamefont
  {Delos}(1988{\natexlab{b}})}]{Du88b}%
  \BibitemOpen
  \bibfield  {author} {\bibinfo {author} {\bibfnamefont {M.~L.}\ \bibnamefont
  {Du}}\ and\ \bibinfo {author} {\bibfnamefont {J.~B.}\ \bibnamefont {Delos}},\
  }\href@noop {} {\bibfield  {journal} {\bibinfo  {journal} {Phys.~Rev.~A}\
  }\textbf {\bibinfo {volume} {38}},\ \bibinfo {pages} {1913} (\bibinfo {year}
  {1988}{\natexlab{b}})}\BibitemShut {NoStop}%
\bibitem [{\citenamefont {Li}\ and\ \citenamefont
  {Tomsovic}(2017{\natexlab{a}})}]{Li17a}%
  \BibitemOpen
  \bibfield  {author} {\bibinfo {author} {\bibfnamefont {J.}~\bibnamefont
  {Li}}\ and\ \bibinfo {author} {\bibfnamefont {S.}~\bibnamefont {Tomsovic}},\
  }\href@noop {} {\bibfield  {journal} {\bibinfo  {journal} {Phys.~Rev.~E}\
  }\textbf {\bibinfo {volume} {95}},\ \bibinfo {pages} {062224} (\bibinfo
  {year} {2017}{\natexlab{a}})},\ \bibinfo {note} {arXiv:1703.07045
  [nlin.CD]}\BibitemShut {NoStop}%
\bibitem [{\citenamefont {Li}\ and\ \citenamefont {Tomsovic}(2018)}]{Li18}%
  \BibitemOpen
  \bibfield  {author} {\bibinfo {author} {\bibfnamefont {J.}~\bibnamefont
  {Li}}\ and\ \bibinfo {author} {\bibfnamefont {S.}~\bibnamefont {Tomsovic}},\
  }\href@noop {} {\bibfield  {journal} {\bibinfo  {journal} {Phys.~Rev.~E}\
  }\textbf {\bibinfo {volume} {97}},\ \bibinfo {pages} {022216} (\bibinfo
  {year} {2018})},\ \bibinfo {note} {arXiv:1712.05568 [nlin.CD]}\BibitemShut
  {NoStop}%
\bibitem [{\citenamefont {Li}\ and\ \citenamefont
  {Tomsovic}(2017{\natexlab{b}})}]{Li17}%
  \BibitemOpen
  \bibfield  {author} {\bibinfo {author} {\bibfnamefont {J.}~\bibnamefont
  {Li}}\ and\ \bibinfo {author} {\bibfnamefont {S.}~\bibnamefont {Tomsovic}},\
  }\href@noop {} {\bibfield  {journal} {\bibinfo  {journal} {J.~Phys.~A:
  Math.~Theor.}\ }\textbf {\bibinfo {volume} {50}},\ \bibinfo {pages} {135101}
  (\bibinfo {year} {2017}{\natexlab{b}})},\ \bibinfo {note} {arXiv:1507.06455
  [nlin.CD]}\BibitemShut {NoStop}%
\bibitem [{\citenamefont {Krauskopf}\ and\ \citenamefont
  {Osinga}(1998)}]{Krauskopf98b}%
  \BibitemOpen
  \bibfield  {author} {\bibinfo {author} {\bibfnamefont {B.}~\bibnamefont
  {Krauskopf}}\ and\ \bibinfo {author} {\bibfnamefont {H.~M.}\ \bibnamefont
  {Osinga}},\ }\href@noop {} {\bibfield  {journal} {\bibinfo  {journal}
  {J.~Comput.~Phys.}\ }\textbf {\bibinfo {volume} {146}},\ \bibinfo {pages}
  {404} (\bibinfo {year} {1998})}\BibitemShut {NoStop}%
\bibitem [{\citenamefont {Mancho}\ \emph {et~al.}(2003)\citenamefont {Mancho},
  \citenamefont {Small}, \citenamefont {Wiggins},\ and\ \citenamefont
  {Ide}}]{Mancho03}%
  \BibitemOpen
  \bibfield  {author} {\bibinfo {author} {\bibfnamefont {A.~M.}\ \bibnamefont
  {Mancho}}, \bibinfo {author} {\bibfnamefont {D.}~\bibnamefont {Small}},
  \bibinfo {author} {\bibfnamefont {S.}~\bibnamefont {Wiggins}}, \ and\
  \bibinfo {author} {\bibfnamefont {K.}~\bibnamefont {Ide}},\ }\href@noop {}
  {\bibfield  {journal} {\bibinfo  {journal} {Physica}\ }\textbf {\bibinfo
  {volume} {D 182}},\ \bibinfo {pages} {188} (\bibinfo {year}
  {2003})}\BibitemShut {NoStop}%
\bibitem [{\citenamefont {Krauskopf}\ \emph {et~al.}(2005)\citenamefont
  {Krauskopf}, \citenamefont {Osinga}, \citenamefont {Doedel}, \citenamefont
  {Henderson}, \citenamefont {Guckenheimer}, \citenamefont {Vladimirsky},
  \citenamefont {Dellnitz},\ and\ \citenamefont {Junge}}]{Krauskopf05}%
  \BibitemOpen
  \bibfield  {author} {\bibinfo {author} {\bibfnamefont {B.}~\bibnamefont
  {Krauskopf}}, \bibinfo {author} {\bibfnamefont {H.~M.}\ \bibnamefont
  {Osinga}}, \bibinfo {author} {\bibfnamefont {E.~J.}\ \bibnamefont {Doedel}},
  \bibinfo {author} {\bibfnamefont {M.~E.}\ \bibnamefont {Henderson}}, \bibinfo
  {author} {\bibfnamefont {J.}~\bibnamefont {Guckenheimer}}, \bibinfo {author}
  {\bibfnamefont {A.}~\bibnamefont {Vladimirsky}}, \bibinfo {author}
  {\bibfnamefont {M.}~\bibnamefont {Dellnitz}}, \ and\ \bibinfo {author}
  {\bibfnamefont {O.}~\bibnamefont {Junge}},\ }\href@noop {} {\bibfield
  {journal} {\bibinfo  {journal} {Int.~J.~Bifurcation~and~Chaos}\ }\textbf
  {\bibinfo {volume} {15}},\ \bibinfo {pages} {763} (\bibinfo {year}
  {2005})}\BibitemShut {NoStop}%
\bibitem [{\citenamefont {Smale}(1963)}]{Smale63}%
  \BibitemOpen
  \bibfield  {author} {\bibinfo {author} {\bibfnamefont {S.}~\bibnamefont
  {Smale}},\ }\href@noop {} {\emph {\bibinfo {title} {Differential and
  Combinatorial Topology}}},\ edited by\ \bibinfo {editor} {\bibfnamefont
  {S.~S.}\ \bibnamefont {Cairns}}\ (\bibinfo  {publisher} {Princeton University
  Press},\ \bibinfo {address} {Princeton},\ \bibinfo {year} {1963})\BibitemShut
  {NoStop}%
\bibitem [{\citenamefont {Smale}(1980)}]{Smale80}%
  \BibitemOpen
  \bibfield  {author} {\bibinfo {author} {\bibfnamefont {S.}~\bibnamefont
  {Smale}},\ }\href@noop {} {\emph {\bibinfo {title} {The Mathematics of Time:
  Essays on Dynamical Systems, Economic Processes and Related Topics}}}\
  (\bibinfo  {publisher} {Springer-Verlag},\ \bibinfo {address} {New York,
  Heidelberg, Berlin},\ \bibinfo {year} {1980})\BibitemShut {NoStop}%
\bibitem [{\citenamefont {Poincar\'e}(1899)}]{Poincare99}%
  \BibitemOpen
  \bibfield  {author} {\bibinfo {author} {\bibfnamefont {H.}~\bibnamefont
  {Poincar\'e}},\ }\href@noop {} {\emph {\bibinfo {title} {Les m\'ethodes
  nouvelles de la m\'ecanique c\'eleste}}},\ Vol.~\bibinfo {volume} {3}\
  (\bibinfo  {publisher} {Gauthier-Villars et fils},\ \bibinfo {address}
  {Paris},\ \bibinfo {year} {1899})\BibitemShut {NoStop}%
\bibitem [{\citenamefont {Easton}(1986)}]{Easton86}%
  \BibitemOpen
  \bibfield  {author} {\bibinfo {author} {\bibfnamefont {R.~W.}\ \bibnamefont
  {Easton}},\ }\href@noop {} {\bibfield  {journal} {\bibinfo  {journal}
  {Trans.~Am.~Math.~Soc.}\ }\textbf {\bibinfo {volume} {294}},\ \bibinfo
  {pages} {719} (\bibinfo {year} {1986})}\BibitemShut {NoStop}%
\bibitem [{\citenamefont {Rom-Kedar}(1990)}]{Rom-Kedar90}%
  \BibitemOpen
  \bibfield  {author} {\bibinfo {author} {\bibfnamefont {V.}~\bibnamefont
  {Rom-Kedar}},\ }\href@noop {} {\bibfield  {journal} {\bibinfo  {journal}
  {Physica~D}\ }\textbf {\bibinfo {volume} {43}},\ \bibinfo {pages} {229}
  (\bibinfo {year} {1990})}\BibitemShut {NoStop}%
\bibitem [{\citenamefont {Bowen}(1975)}]{Bowen75}%
  \BibitemOpen
  \bibfield  {author} {\bibinfo {author} {\bibfnamefont {R.}~\bibnamefont
  {Bowen}},\ }\href@noop {} {\emph {\bibinfo {title} {Lect. Notes in Math. Vol.
  470.}}}\ (\bibinfo  {publisher} {Springer-Verlag},\ \bibinfo {address}
  {Berlin},\ \bibinfo {year} {1975})\BibitemShut {NoStop}%
\bibitem [{\citenamefont {Gaspard}(1998)}]{Gaspard98}%
  \BibitemOpen
  \bibfield  {author} {\bibinfo {author} {\bibfnamefont {P.}~\bibnamefont
  {Gaspard}},\ }\href@noop {} {\emph {\bibinfo {title} {Chaos, Scattering and
  Statistical Mechanics}}}\ (\bibinfo  {publisher} {Cambridge University
  Press},\ \bibinfo {address} {Cambridge, UK},\ \bibinfo {year}
  {1998})\BibitemShut {NoStop}%
\bibitem [{\citenamefont {Hadamard}(1898)}]{Hadamard1898}%
  \BibitemOpen
  \bibfield  {author} {\bibinfo {author} {\bibfnamefont {J.}~\bibnamefont
  {Hadamard}},\ }\href@noop {} {\bibfield  {journal} {\bibinfo  {journal}
  {J.~Math.~Pures~Appl.~series 5}\ }\textbf {\bibinfo {volume} {4}},\ \bibinfo
  {pages} {27} (\bibinfo {year} {1898})}\BibitemShut {NoStop}%
\bibitem [{\citenamefont {Birkhoff}(1927{\natexlab{a}})}]{Birkhoff27a}%
  \BibitemOpen
  \bibfield  {author} {\bibinfo {author} {\bibfnamefont {G.~D.}\ \bibnamefont
  {Birkhoff}},\ }\href@noop {} {\emph {\bibinfo {title} {A.M.S. Coll.
  Publications, vol. 9}}}\ (\bibinfo  {publisher} {American Mathematical
  Society},\ \bibinfo {address} {Providence},\ \bibinfo {year}
  {1927})\BibitemShut {NoStop}%
\bibitem [{\citenamefont {Birkhoff}(1935)}]{Birkhoff35}%
  \BibitemOpen
  \bibfield  {author} {\bibinfo {author} {\bibfnamefont {G.~D.}\ \bibnamefont
  {Birkhoff}},\ }\href@noop {} {\bibfield  {journal} {\bibinfo  {journal}
  {Mem.~Pont.~Acad.~Sci.~Novi~Lyncaei}\ }\textbf {\bibinfo {volume} {1}},\
  \bibinfo {pages} {85} (\bibinfo {year} {1935})}\BibitemShut {NoStop}%
\bibitem [{\citenamefont {Morse}\ and\ \citenamefont
  {Hedlund}(1938)}]{Morse38}%
  \BibitemOpen
  \bibfield  {author} {\bibinfo {author} {\bibfnamefont {M.}~\bibnamefont
  {Morse}}\ and\ \bibinfo {author} {\bibfnamefont {G.~A.}\ \bibnamefont
  {Hedlund}},\ }\href@noop {} {\bibfield  {journal} {\bibinfo  {journal}
  {Amer.~J.~Math.}\ }\textbf {\bibinfo {volume} {60}},\ \bibinfo {pages} {815}
  (\bibinfo {year} {1938})}\BibitemShut {NoStop}%
\bibitem [{\citenamefont {H\'enon}(1976)}]{Henon76}%
  \BibitemOpen
  \bibfield  {author} {\bibinfo {author} {\bibfnamefont {M.}~\bibnamefont
  {H\'enon}},\ }\href@noop {} {\bibfield  {journal} {\bibinfo  {journal}
  {Comm.~Math.~Phys.}\ }\textbf {\bibinfo {volume} {50}},\ \bibinfo {pages}
  {69} (\bibinfo {year} {1976})}\BibitemShut {NoStop}%
\bibitem [{\citenamefont {Cvitanovi\'{c}}\ \emph {et~al.}(1988)\citenamefont
  {Cvitanovi\'{c}}, \citenamefont {Gunaratne},\ and\ \citenamefont
  {Procaccia}}]{Cvitanovic88a}%
  \BibitemOpen
  \bibfield  {author} {\bibinfo {author} {\bibfnamefont {P.}~\bibnamefont
  {Cvitanovi\'{c}}}, \bibinfo {author} {\bibfnamefont {G.}~\bibnamefont
  {Gunaratne}}, \ and\ \bibinfo {author} {\bibfnamefont {I.}~\bibnamefont
  {Procaccia}},\ }\href@noop {} {\bibfield  {journal} {\bibinfo  {journal}
  {Phys.~Rev.~A}\ }\textbf {\bibinfo {volume} {38}},\ \bibinfo {pages} {1503}
  (\bibinfo {year} {1988})}\BibitemShut {NoStop}%
\bibitem [{\citenamefont {Cvitanovi\'{c}}(1991)}]{Cvitanovic91}%
  \BibitemOpen
  \bibfield  {author} {\bibinfo {author} {\bibfnamefont {P.}~\bibnamefont
  {Cvitanovi\'{c}}},\ }\href@noop {} {\bibfield  {journal} {\bibinfo  {journal}
  {Physica~D}\ }\textbf {\bibinfo {volume} {51}},\ \bibinfo {pages} {138}
  (\bibinfo {year} {1991})}\BibitemShut {NoStop}%
\bibitem [{\citenamefont {Hagiwara}\ and\ \citenamefont
  {Shudo}(2004)}]{Hagiwara04}%
  \BibitemOpen
  \bibfield  {author} {\bibinfo {author} {\bibfnamefont {R.}~\bibnamefont
  {Hagiwara}}\ and\ \bibinfo {author} {\bibfnamefont {A.}~\bibnamefont
  {Shudo}},\ }\href@noop {} {\bibfield  {journal} {\bibinfo  {journal}
  {J.~Phys.~A: Math.~Gen.}\ }\textbf {\bibinfo {volume} {37}},\ \bibinfo
  {pages} {10521–10543} (\bibinfo {year} {2004})}\BibitemShut {NoStop}%
\bibitem [{\citenamefont {Birkhoff}(1927{\natexlab{b}})}]{Birkhoff27}%
  \BibitemOpen
  \bibfield  {author} {\bibinfo {author} {\bibfnamefont {G.~D.}\ \bibnamefont
  {Birkhoff}},\ }\href@noop {} {\bibfield  {journal} {\bibinfo  {journal} {Acta
  Math.}\ }\textbf {\bibinfo {volume} {50}},\ \bibinfo {pages} {359} (\bibinfo
  {year} {1927}{\natexlab{b}})}\BibitemShut {NoStop}%
\bibitem [{\citenamefont {Moser}(1956)}]{Moser56}%
  \BibitemOpen
  \bibfield  {author} {\bibinfo {author} {\bibfnamefont {J.}~\bibnamefont
  {Moser}},\ }\href@noop {} {\bibfield  {journal} {\bibinfo  {journal}
  {Commun.~Pure Appl.~Math.}\ }\textbf {\bibinfo {volume} {9}},\ \bibinfo
  {pages} {673} (\bibinfo {year} {1956})}\BibitemShut {NoStop}%
\bibitem [{\citenamefont {da~Silva~Ritter}\ \emph {et~al.}(1987)\citenamefont
  {da~Silva~Ritter}, \citenamefont {Ozorio~de Almeida},\ and\ \citenamefont
  {Douady}}]{Silva87}%
  \BibitemOpen
  \bibfield  {author} {\bibinfo {author} {\bibfnamefont {G.~L.}\ \bibnamefont
  {da~Silva~Ritter}}, \bibinfo {author} {\bibfnamefont {A.~M.}\ \bibnamefont
  {Ozorio~de Almeida}}, \ and\ \bibinfo {author} {\bibfnamefont
  {R.}~\bibnamefont {Douady}},\ }\href@noop {} {\bibfield  {journal} {\bibinfo
  {journal} {Physica~D}\ }\textbf {\bibinfo {volume} {29}},\ \bibinfo {pages}
  {181} (\bibinfo {year} {1987})}\BibitemShut {NoStop}%
\bibitem [{\citenamefont {Harsoula}\ \emph {et~al.}(2015)\citenamefont
  {Harsoula}, \citenamefont {Contopoulos},\ and\ \citenamefont
  {Efthymiopoulos}}]{Harsoula15}%
  \BibitemOpen
  \bibfield  {author} {\bibinfo {author} {\bibfnamefont {M.}~\bibnamefont
  {Harsoula}}, \bibinfo {author} {\bibfnamefont {G.}~\bibnamefont
  {Contopoulos}}, \ and\ \bibinfo {author} {\bibfnamefont {C.}~\bibnamefont
  {Efthymiopoulos}},\ }\href@noop {} {\bibfield  {journal} {\bibinfo  {journal}
  {J.~Phys.~A: Math.~Theor.}\ }\textbf {\bibinfo {volume} {48}},\ \bibinfo
  {pages} {135102} (\bibinfo {year} {2015})},\ \bibinfo {note}
  {arXiv:1502.00664 [nlin.CD]}\BibitemShut {NoStop}%
\bibitem [{\citenamefont {Mitchell}\ \emph {et~al.}(2003)\citenamefont
  {Mitchell}, \citenamefont {Handley}, \citenamefont {Tighe}, \citenamefont
  {Delos},\ and\ \citenamefont {Knudson}}]{Mitchell03a}%
  \BibitemOpen
  \bibfield  {author} {\bibinfo {author} {\bibfnamefont {K.~A.}\ \bibnamefont
  {Mitchell}}, \bibinfo {author} {\bibfnamefont {J.~P.}\ \bibnamefont
  {Handley}}, \bibinfo {author} {\bibfnamefont {B.}~\bibnamefont {Tighe}},
  \bibinfo {author} {\bibfnamefont {J.~B.}\ \bibnamefont {Delos}}, \ and\
  \bibinfo {author} {\bibfnamefont {S.~K.}\ \bibnamefont {Knudson}},\
  }\href@noop {} {\bibfield  {journal} {\bibinfo  {journal} {Chaos}\ }\textbf
  {\bibinfo {volume} {13}},\ \bibinfo {pages} {880} (\bibinfo {year}
  {2003})}\BibitemShut {NoStop}%
\end{thebibliography}%

\end{document}